# Raman and ATR FT-IR investigations of innovative silica nanocontaniers loaded with a biocide for stone conservation treatments

L. Ruggiero,[1] A. Sodo,[1] M. Cestelli-Guidi,[2] M. Romani,[2] A. Sarra[1,3], P. Postorino[3] and M.A. Ricci[1]

[1] *Dipartimento di Scienze, Università degli Studi "Roma Tre", Via della Vasca Navale, 86-00146 Roma, Italia*

[2] *INFN-Laboratori Nazionali di Frascati, Via Enrico Fermi, 40 - 00044 Frascati (RM), Italia*

[3] *Dipartimento di Fisica, Università degli Studi "Sapienza", Piazzale Aldo Moro, Roma, Italia*

**Abstract**

In the last years, significant research efforts have been devoted to the reduction of environmental impact of biocides and to the improvement of their effectiveness over time against degradation of stone artifacts. This work reports the spectroscopic characterization of two silica nanosystems, with different geometry and structure, loaded with a biologically-active compound, the 2-mercaptobenzothizole (MBT), for controlled antifouling release on the surface of outdoor stone artefacts.

We have combined FT-IR spectroscopy in ATR mode and Raman spectroscopy to investigate the biocide loading mechanism in both nanosystems from the physical and chemical point of view. These techniques demonstrate that the geometry of confinement influences the loading of the nanoparticles and their interaction with the confining medium, with possible consequences in the biocide release rate. Moreover we have seen that the biocide interacts with the surfactant (cetrimonium bromide, CTAB) and tends to dimerize.

**Introduction**

Most of the worldwide Cultural Heritage monuments are built using stone material. Different types of natural and man-made stone materials (concrete, brickwork, mortar) have extremely variable physical and chemical properties (texture, high porosity etc), resulting with widely different abilities to resist weathering (durability)[1]. Decay of stone materials is due to their interaction with the environment: weathering and atmospheric conditions (such as light, temperature, humidity, pollutants and acid rain) [2-5]. The most immediate consequence of this interaction is a chemical and physical alteration followed, in most cases, by biological colonization. The biodeterioration of stone is associated to nearly all environmentally induced degradation processes in a synergic manner, as the presence of one deterioration process enhances the effectiveness of the other [6-8].

Currently, to reduce the biological colonization on outdoor surfaces, the biocides are applied directly on the stone surface or added into coating formulations [9-10]. These methods, however, have several important drawbacks. When the biocides are directly applied on the stone surface the treatment is effective for a limited period of time and the biocides easily deteriorate and/or are washed away. When small biocides molecules are added into the coating formulation, are subject to fast self-diffusion in the coating matrix [11-12].

These disadvantages can be smartly faced by the encapsulation of biocides in nanocontainers and by uniformly dispersing them in the coating. The dispersed nanocontainers can release the biocide on demand, during the life of the applied coating. Literature [12-14] reports the synthesis and characterization of a few systems, such as mesoporous silica-based materials, for controlled release of biocides over time. In particular, in our laboratory, we synthesized two different silica-based nanocontainers, namely a core-shell nanocapsule (Si-NC) and a mesoporous nanoparticles (Si-MNP) [15-16]. Both have been loaded with 2-mercaptobenzothiazole (MBT), one of the most common commercial biocide [16], in order to preclude the direct contact of the biocide with the environment. Both synthesis procedures are one-step self-assembly methods, involving the tetraethyl orthosilicate (TEOS) polymerization assisted by a cationic surfactant (Cetrimonium bromide, CTAB) and the biocide. According to literature, the two synthetic procedures physically confine the thiazole derivative, without any chemical interaction. On the other hand, possible chemical interaction of the synthesis reagents and the bioactive compound (MBT) are expected to influence the release processes. In this work, we try to investigate the nature of the confinement of the MBT into the two different silica nanocontainers, namely (Si-NC) and (Si-MNP), by FT-IR and Raman spectroscopy. Both these techniques are very useful to analyze the influence of the nano - structure on the molecular mobility of the guest molecules in the nanosized pores and channels.
In addition, we report ζ-potential comparative measurements, to highlight the different confinement of the biocide in the two different silica structures.

**Materials and methods**

**Materials**

The investigated materials are two different kinds of silica nanocontainers synthesized according the methods reported, in details, in a previous work [16].

The first synthesis procedure is an encapsulation procedure: the cationic surfactant (CTAB), in a aqueous basic environment, forms minimicelles. The biocide is added during the synthesis using the diethyl ether as cosolvent and oil phase in the oil-in-water miniemulsion. The silica precursor (TEOS) condensate at the oil-in-water miniemulsion interface and during this step the diethyl ether evaporates producing the gradual mesoporosity of the shell, thus generating the core-shell nanocapsule (Si-NC).

In the second synthesis procedure, after the micellization of the cationic surfactant in basic aqueous environment, the biocide is added in the synthesis bath with organic solvent. The hydrolysis and condensation of silica precursors in basic environment at 80 ˚C is then employed for the formation of silica nanoparticles with hexagonally packed mesopores (Si-MNP).

The silica nanocapsules (Si-NC) are spherical particles with a regular shape and a hollow nature. The Si-NC loaded with the MBT have a narrow size distribution centred at 133 nm with a standard deviation of 33 nm (estimated by 90 measurements) and a surface area of 720 $m^2g^{-1}$ and a pore volume of 1.62 $cm^3g^{-1}$, as reported in our previous work [16].

The nanoparticles (Si-MNP) loaded with MBT show a normal distribution centred at 40 nm with a surface area of 924 $m^2g^{-1}$ and a pore volume of 1.86 $cm^3g^{-1}$ [16].

The biocide, encapsulated in these two types of nanocontainers is isolated from the external environment through a physical barrier of mesoporous silica. The presence of the biocide in the system and its interaction with the silica network can be probed by means of complementary spectroscopic techniques, specifically FTIR and Raman.

The MBT is an organosulfur compound and the molecule consists of a benzene ring fused to a 2-mercaptothiazole ring, which includes an S atom within the ring structure (endocyclic) and one without (exocyclic) and an N atom. One of the most interesting properties of MBT is its existence of thiol and thione tautomeric form; according to the literature, the thione form is the prevailing specie [17].

**Fourier Transform IR spectroscopy (FTIR)**

The infrared (IR) spectra of the synthesis reagents (e.g. MBT and CTAB) and the nanosilica systems were investigated using a Bruker TensorII FTIR spectrometer in Attenuated Total Reflection (ATR) mode. Samples, in the form of powder, were placed on the top of a diamond ATR crystal and pressed to obtain complete adhesion at the interface. The spectra were measured in the 4000–400 $cm^{-1}$ range by coadding 256 scans at 2 $cm^{-1}$ resolution. Due to the high refractive index of diamond ($n$=2.4) the penetration depth of the IR beam into the sample is about 2 microns, giving spectroscopic information comparable to the one obtained in transmission mode.

**Raman spectroscopy**

Raman measurements have been performed at room temperature using an inVia Renishaw Raman spectrometer equipped with a diode laser (785 nm, output power 200 mW), an edge filter to select the Raman scattering avoiding the elastic contribution , a 1200 lines/mm diffraction grating and a Peltier cooled 1024x256 pixel CCD detector. Samples have been mounted on the manual stage of a Leica DM2700 M confocal microscope. Focusing of the laser beam and collection of Raman signals has been realised by a 50x long-working distance objective. The spectra have been recorded using a laser power at about 39 mW on the sample. Several scans (20) have been collected in order to improve the signal-to-noise ratio with the acquisition time of 10 s. The Raman spectrometer has been calibrated prior to the measurements using a Si wafer and by performing the automatic offset correction. The spectra acquisition and data analyses have been accomplished using WiRE™ and Origin softwares. To get a representative spectrum for each sample, ten spectra, collected on different point, have been averaged and normalized. The peak positions are estimated to be accurate to at least ±2 $cm^{-1}$.

**Zeta Potential Measurements**

Electrophoretic mobility measurements were performed using a Malvern NanoZetaSizer apparatus, equipped with a 5mW HeNe laser (Malvern Instruments Ltd, UK) in a quasi-backscattering configuration (the diffused light is collected at an angle of 173°). Measurements were carried out using Mixed Mode Measurement (M3) method combined with Phase Analysis Light Scattering (PALS) technique, using capillary cuvettes with palladium electrodes (disposable cuvettes, Malvern, UK). The measured electrophoretic mobility $\mu e$ was converted into Zeta potential through the Smoluchowski relation $\zeta = \mu e \eta / \varepsilon$, where $\varepsilon$ is the permittivity of the solvent and $\eta$ is the viscosity.

**Results and discussion**

The two different silica nanoreservoirs systems loaded with 2-mercaptobenzothiazole , along with the empty silica nanosystems prepared in the absence of biocide (MBT), i.e., obtained solely from TEOS and CTAB have analyzed by FT-IR and Raman spectroscopy. The latter samples have been used as control, to evaluate the silica bands, its network vibrations and the residual of the cationic surfactant (CTAB).

The ATR FT-IR spectra of both nanosystems (Si-NC and Si-MNP) show the characteristic bands associated to the silica shell. Silica presents a characteristic region of peaks from 1250 to 700 $cm^{-1}$ that can provide structural characteristics of the network. Specifically, the large band centered at 1085 $cm^{-1}$, corresponding to the asymmetric $\nu(Si–O–H)$ stretching mode, can be deconvoluted in four components, two longitudinal (LO) and two transverse (TO) optic modes, related to the different arrangement of siloxane rings, namely the four-membered $(SiO)_4$ and the six-membered $(SiO)_6$ arrangement. Literature reports that the position and relative intensities of the LO and TO modes are shifted with the introduction of chemical groups or organic molecules in the silica network [18].

When the biocide is encapsulated in to the Si-NCs, the presence of MBT is confirmed by the peak at 754 $cm^{-1}$ due to the C-S- stretching. Moreover we observe a broadening and a shift of the Si-O-Si asymmetric stretching band frequency towards higher wavelengths. The second derivative of the absorbance spectrum (Fig. 1a) enhances the band shift.

Shifts to higher wave numbers are related to the network deformation needed to accommodate the MBT organic groups within the inorganic silica matrix, resulting in larger siloxane rings and greater Si–O–Si angles and longer Si–O bond lengths. This may be the case for Si-NP and not for Si-MNP, where the biocide is placed in the interstitial space among silica network.
Other spectral variations are observed in the region of the C=S and C-C stretching modes of the biocide that are superimposed to the silica signal. The same applies to the peak at 1228 $cm^{-1}$ that shifts of 10 $cm^{-1}$ because of the strong band of the biocide at 1243 $cm^{-1}$ attributed to the C-N stretching.

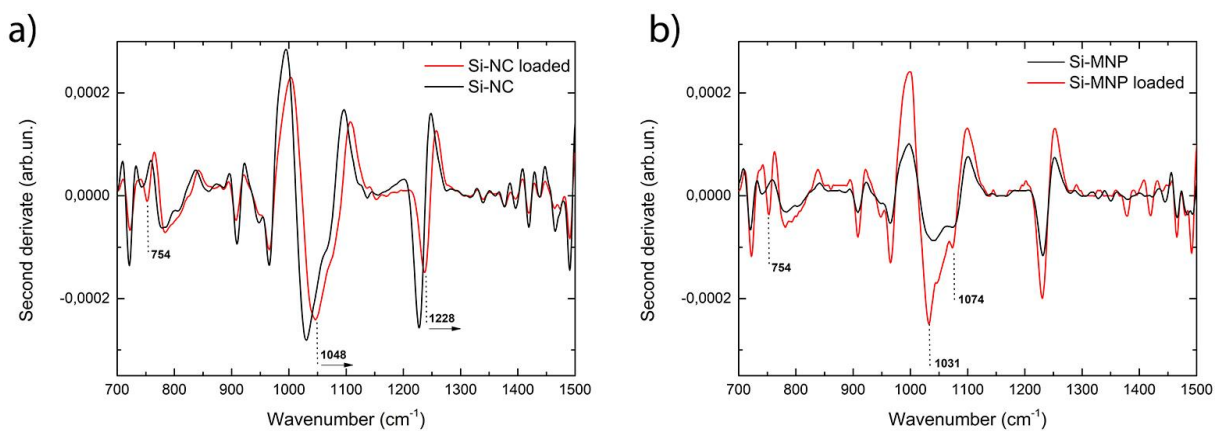

*Fig.1 ATR FT-IR spectra second derivate of the Si-NC (a) and Si-MNP (b) loaded (red) and empty (black).*

When the biocide is entrapped in the Si-MNPs, the ATR FT- IR spectrum shows, in addition to the peak at 754 cm$^{-1}$, two well-defined bands at 1031 and 1074 cm$^{-1}$, which are associated with the C=S stretching and SH bending modes of the MBT (Fig.1b). According to the literature, these data confirm the presence of the biocide in the powders under analysis and underline that the biocide molecules are subject to a different spatial confinement, compared to the Si-NCs case.

To better confirm the hypothesis of different interaction of the biocide with the silica structure in the two synthesized compounds, structural information will be provided, as well as *in-situ* analysis to follow the structural network arrangement during the synthesis procedures.

Raman spectra of the two nanosystems empty and loaded are reported in Fig. 2 and Fig. 3a, respectively. Here we discuss only the 100–1,800 cm$^{-1}$ region of the spectra: this is indeed the fingerprint region of the loaded nanosystems, which may be sensible to the presence of the biocide (MBT). The detailed assignment of all the bands of the biocide, the cationic surfactant and the silica structure reported in the literature [19-24], will guide our interpretation of the spectra of the two nanosystems.

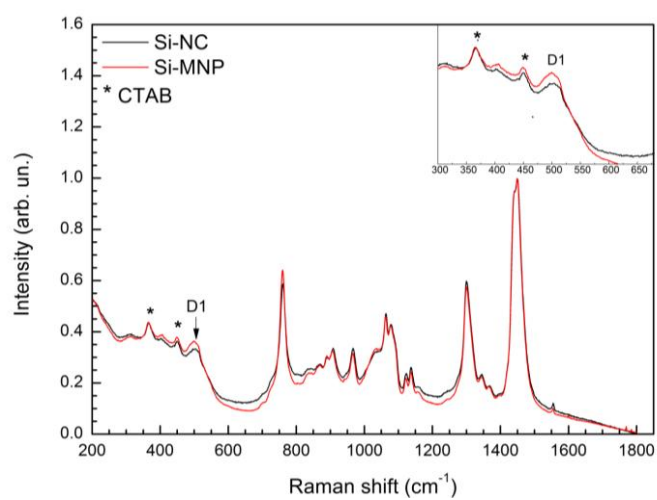

*Fig 2. Raman spectra of the silica nanocapsules (black) and the silica nanoparticles (red).*

In particular, the investigated region provides information on the structural characteristics of the silica nanocontainers and gives information about the degree of polymerization of the silica precursor [18]. The Raman measurements highlight that the polymerization of the precursor is completed and this information is confirmed by the absence of the stretching mode ν(Si-OH), at 945 cm$^{-1}$, typically assigned to the hydrophilic residual silanol group. [19-20] In the range between 200 and 650 cm$^{-1}$, the only easily detectable vibrational contribution from the containers is the D1 band at 495 cm$^{-1}$, due to the breathing mode of the siloxane ring (with 3 or 4 SiO units). Previous studies report that the amplitude of this band gives qualitative information about the specific surface and the dimensions of the nanosystems, as it increases with the specific surface area [21-22]. In our case, the amplitude of this band is larger in Si-MNP compared to Si-NC, according to the surface area data (see inset in Fig.2). Other vibrational features ascribable to silica structure are not clearly identifiable because of the presence of the CTAB contribution.

The vibrational contributions of the CTAB enrich the spectra of the nanosystems (both empty and loaded) with many strong bands, as summarized in Table 1. CTAB plays in both synthesis the role of templating agent. Although after each synthesis, the final product was washed with water several times to remove the CTAB, yet the Raman and FT-IR spectra evidence the presence of its residues. Interestingly the ratio of CTAB to silica seems the same for the two samples, as suggested by the intensity ratio of the CTAB peaks after the spectra normalization: this proves the good reproducibility of the synthesis protocol.

The encapsulation of the 2-mercaptobenzothiazole into the silica nanocontainers is confirmed also by the Raman measurements, reported in Fig.3a.

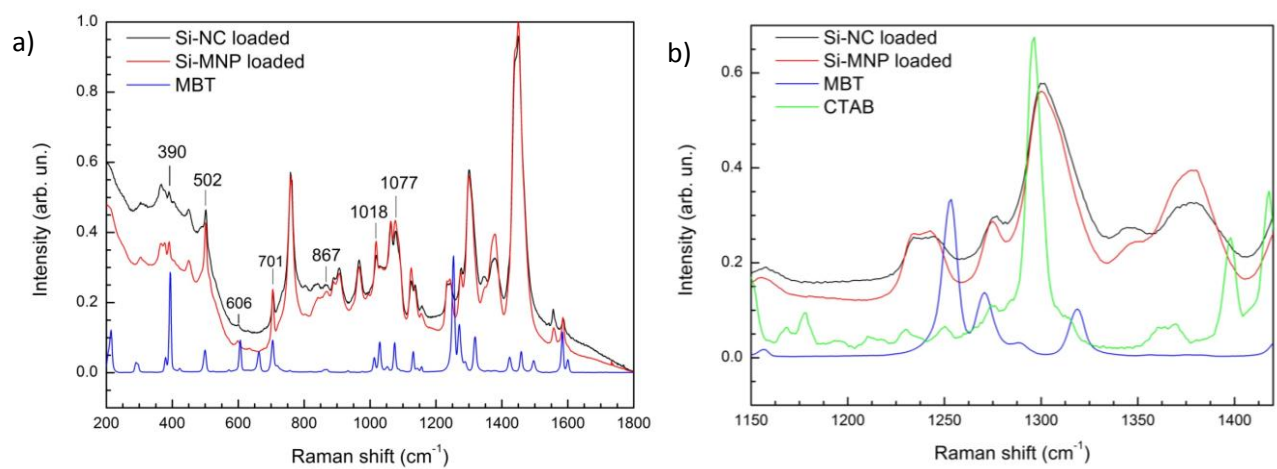

*Fig.3 a) Raman spectra of both silica nanocontainers loaded with MBT, compared with MBT standard spectrum; b) details of the Raman spectra reported in a) compared with the spectra of the pure MBT and CTAB powders.*

At low wavenumbers, the peaks that can be assigned to the biocide are superimposed on the spectral profile of the silica signals. The well-defined band observed at 390 cm$^{-1}$ is ascribed to S-C=S bending vibration. The band at 502 cm$^{-1}$ is ascribable to CCC and CCN bending. The stretching of C-S bond is evident at 606 and 701 cm$^{-1}$. At 867 cm$^{-1}$ we can observe CCC bending, CC and CN stretching. In the region between 1000-1100 cm$^{-1}$ it is possible to discriminate the vibrational contribution of the endocyclic and exocyclic sulphur atoms, respectively: the band at 1018 cm$^{-1}$ is associated to CH bending, CS stretching in S-C-S and the band at 1077 cm$^{-1}$ is related to CS stretching in S-C-S, CCC bending, C=S stretching [17].

Interestingly, in addition to the above quoted bands, new bands, absent in the reference spectra, are observed in the regions between 1230-1250 cm$^{-1}$ and 1330-1410 cm$^{-1}$ (see Fig.3b).

The shorter wavenumber range is typically assigned to the vibrational feature of the CN stretching of the biocide and the CN stretching and CNC bending of CTAB. The spectra of the loaded nanosystems, in fact, compared to the pure compound, show a splitting of the band at 1253 and a red shift of the peak at 1270 cm$^{-1}$ [23-24]. These features may be the signature of a chemical reaction between CTAB and MBT [25-26].

The longer wavenumber range (1330-1410 cm$^{-1}$), instead, suggests that the MBT, in both the synthesis conditions, is more stable in the dimeric conformation. Indeed according to the literature, the bands observed in this spectral region can be attributed to the CH and NH bending, and the CN stretching of the dimeric complex [27], formed by two molecules in thione form linked by linear N-H⋯S hydrogen bonds [28].

All the assignments are summarised in Table 1.

Tab.1 *Vibrational wavenumbers in cm$^{-1}$ of silica nanocapsules (Si-NC) loaded and empty, and silica nanoparticles (Si-MNP) loaded and empty, compared with the vibrational wavenumbers of the biocide (MBT) and the cationic surfactant (CTAB). In bold the vibrational assignments related to the presence of the MBT.*

| Si_NC | Si_NC loaded | Si_MNP loaded | Si_MNP | MBT | CTAB |
|---|---|---|---|---|---|
| - | **390** | **390** | - | **394** | - |
| 450 | 450 | 450 | 450 | - | 450 |
| 502 | **502** | **502** | 500 | **500** | 500 |
| - | **598** | **603** | 603 | **607** | - |
| - | **701** | **701** | - | **701** | - |
| 761 | 759 | 759 | 758 | 756 | 761 |
| 837 | 838 | 840 | 833 | - | 831 |
| 866 | **867** | **869** | 870 | **869** | 867 |
| 890 | 890 | 890 | 890 | - | 887 |
| 909 | 907 | 907 | 907 | - | 905 |
| 965 | 965 | 965 | 966 | - | 960 |
| - | **1018** | **1019** | - | **1013** | 1011 |
| 1062 | 1062 | 1062 | 1063 | 1053 | 1062 |
| **1080** | **1079** | **1077** | **1077** | **1074** | - |

Summing up, Raman results clarify that the biocide interacts with the cationic surfactant through the nitrogen atom and tends to dimerize.
Moreover, FT-IR and Raman measurements show that the biocide contributions are more intense in the mesoporous nanoparticles (Si-MNP). This suggests that, while in Si-NC the biocide is encapsulated in the core of the silica nanocapsule, it is dispersed into the matrix of the mesoporous nanoparticles in the case of Si-MNP, thus being more easily detected. To verify this suggestion, ζ-potential measurements as a function of time have been performed.

Si-NCs and Si-MNPs have been suspended in deionized water and electrophoretic mobility has been measured as a function of time starting from immediately after the suspension preparation (t =

0 *s*). At each time we obtained a rather narrow Gaussian distribution of ζ-potential values, that is about 5 mV / 7 mV for empty/loaded Si-NC and 4 mV / 5 mV for empty/loaded Si-MNP. The central values of these distributions are shown as a function of time in Figures 4a and 4b, for both empty and MBT loaded nanosystems. Solid and dashed lines represent the best linear fit of the experimental data and best fit parameters are reported in the legends.

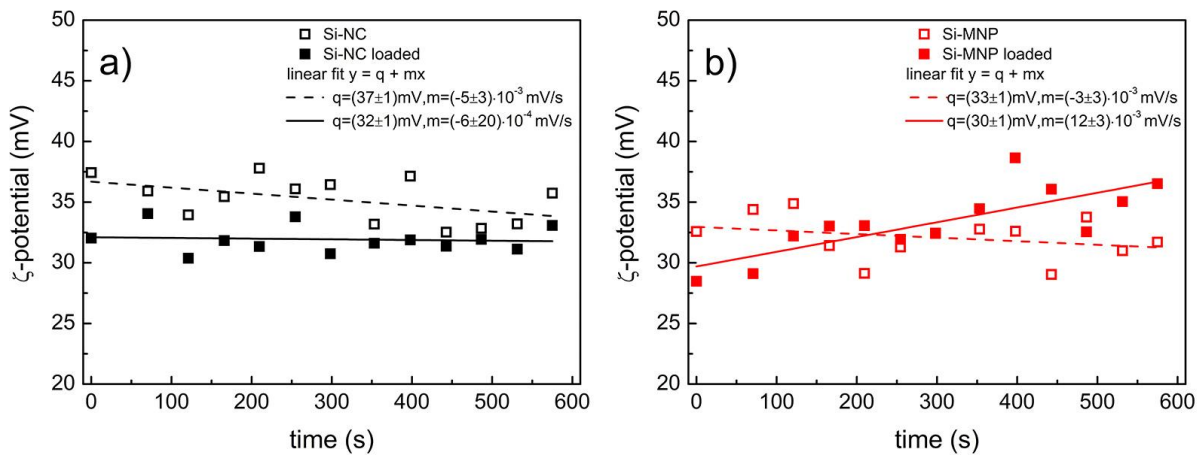

*Fig. 4 Central values of the ζ-potential distributions measured as a function of time: a) empty (black open squares) and MBT loaded (black filled squares) Si-NC; b) empty (red open squares) and MBT loaded (red filled squares) Si-MNP. Best fit lines (dashed for empty and solid for loaded) and relative parameter values are reported in the two panels.*

In the case of Si-NCs (Fig. 4a) the central values of the ζ-potential distribution are roughly constant (the slope best fit values differ from zero by about $10^{-3}$ mV/s) for both empty and loaded samples. The intercept values, thus actually the average ζ-potential values, provide (37±1) mV and (32±1) mV for empty and loaded systems respectively. It is worth to notice that in the latter case a smaller value is measured as expected in presence of negatively charged MBT in aqueous environment.

For Si-MNPs (Fig. 4b) we observed a similar difference between empty and loaded samples at t=0 *s* i.e. (33±1) mV and (30±1) mV respectively. While for empty Si-MNP the ζ-potential central values are basically time-independent, the loaded sample shows a slight but significant increase starting from the suspension preparation time (t = 0*s*). This behaviour is compatible with a progressive MBT release as well as with a greater exposure of MBT biocide in Si-MNP with respect to Si-NC.

ζ-potential results, although preliminary, seem to confirm the different confinement geometry and interaction of the biocide in both synthesized systems, as suggested by Raman spectroscopy.

**Conclusion**

The present paper reports the spectroscopic investigations of two different silica nanosystems developed for the encapsulation of 2-mercaptobenzothiazole. FT-IR and Raman results confirm that our protocol is successful as far as the loading of the biocide is concerned, and suggest that the biocide chemically interacts with the surfactant CTAB and not with the confining silica matrix. The absence of chemical bonds with silica is a relevant result, as guarantees the molecular mobility needed to exploit the antifouling activity of the product. The chemical interaction with CTAB can instead explain why we were not able to completely wash it away. Moreover Raman spectra and ζ-potential measurements evidence differences between Si-NC and Si-MNP loaded particles, suggesting that the biocide is confined within the Si-NC internal volume, and embedded within the sponge-like structure of the Si-MNP particles. This implies that when the two nanocontainers will be dispersed in a medium, as for instance a protective coating for stone materials, will release the biocide with quite different rate[16].

**Acknowledgment**

This project is financed by SUPERARE grant "Gruppi di Ricerca", Regione Lazio. We thank prof. F. Bordi (Physics Deptartment, Sapienza University) for the access on Malvern NanoZetaSizer instrument, for Zeta potential measurements.